# "Single-Atom" Catalysis: An Opportunity For Surface Science

*Gareth S. Parkinson*

Institute of Applied Physics, TU Wien, Vienna, Austria

*Corresponding Author: Gareth.parkinson@tuwien.ac.at

ABSTRACT

Over the past decade, extensive research into "single-atom" catalysts (SACs) has revealed that the catalytic behavior of metal adatoms is highly dependent on how they interact with their support. A strong dependence on the local coordination environment has led to comparisons with metal-organic complexes, and there is growing excitement about the potential to fine-tune SACs by controlling the adsorption geometry. The rise of computational screening to identify the optimal support/metal combinations underscores the need for rigorous benchmarking of theoretical methods, to validate realistic geometries, mechanisms, and the impact of adsorption on stability and catalytic activity. The surface science approach is particularly well-suited for this task because it allows to precisely determine the geometry of the metal atom and interpret its catalytic behavior. Moreover, the effects of temperature and molecular adsorption on the model catalysts stability can be studied in isolation, and conclusions drawn from UHV studies tested in increasingly common near-ambient pressure and electrochemical setups. This perspective highlights recent breakthroughs and specific systems—including metal oxides, metal-organic frameworks, and carbon nitrides—where insights from surface science experiments can significantly advance understanding in this rapidly evolving field.

The field of surface science grew out of a desire to understand what occurs during catalytic reactions. As far back as 1922[1], Langmuir had concluded that the complexity of industrial catalysts would likely preclude unraveling the multitude of effects at play. The solution, common to many branches of science, was to develop model systems that are simple enough to be tractable, but still mimic the real system in a meaningful way. Because metal particles were the active catalytic component, and they were known to expose preferred facets, oriented single-crystal samples became the model of choice. The advent of vacuum technology enabled the preparation and maintenance of clean surfaces, sparking technological developments that allowed rapid advancements in determining surface structure[2] and spectroscopic methods to understand reactant adsorption and catalytic mechanisms. This approach led to significant insights into reactions like CO oxidation (important for automobile emissions)[3] and the Haber-Bosch process (for ammonia production)[4], earning Gerhard Ertl the 2007 Nobel Prize in Chemistry for pivotal contributions to surface science. A detailed summary of his and other early contributions in this field is provided in Reference 5.

As with any model, there are always possibilities for improvement. The terms "pressure gap" and "materials gap" refer to two key limitations: the significant difference between the ultra-high vacuum pressures used in laboratory experiments versus real catalytic conditions[6], and the complexity of real catalysts, which often have active sites absent on a metal single-crystal surface. Bridging the pressure gap has been a longstanding goal in surface science, and today "near-ambient pressure" (NAP) adaptations of many critical techniques—such as surface x-ray diffraction (SXRD), x-ray photoelectron spectroscopy (XPS) and scanning tunneling microscopy (STM)—are well established and increasingly applied[7-11]. Recent trends in electrochemical research have similarly spurred efforts to develop protocols for electrochemical surface science[12]. Efforts to bridge the materials gap are similarly numerous. One important development was the study of metal-oxide surfaces[13], which are common support materials (as well as active catalysts for some reactions). Experiments on $TiO_2(110)$, probably the most studied metal-oxide surface[14], utilized single crystal samples rendered conducting by *in situ* reduction, while epitaxial thin film growth allows to create samples with sufficient conductivity even for insulating materials[15-17]. Once the structure and properties of the metal-oxide support are well understood, the next step in complexity is to add metal nanoparticles. A clear size effect was discovered, and a non-scalable regime where the conclusions derived from prior studies on metal single crystals do not hold[18, 19]. One notable example was the discovery that Au becomes catalytically active for CO oxidation once nanoparticles become smaller than circa. 10 nm in diameter[14]. The importance of the metal-metal oxide interface has been widely studied[20], leading to a focus on "inverse catalysts"—metal oxide islands on metal single-crystal supports[21]. Many concepts derived from these studies are directly relevant to SAC, where all metal atoms are in direct contact with the support.

The development of the size-selected cluster source provided an elegant method to study size effects in catalysis. This technology allows for the deposition of metal particles with a narrow size distribution, enabling single-atom precision in reactivity studies, provided the clusters do not undergo significant sintering on the single-crystal support[22-24]. Many instances were uncovered where "magic" cluster sizes are particularly active due to some coincidence of physical and electronic structure. It was one such study[25] that provided some of the first surface science evidence for SAC. As the subtitle of the paper ("one atom is enough!") suggests, Heiz and coworkers reported the somewhat surprising result that a single Pd atom supported on MgO(001) could catalyze the cyclotrimerization of acetylene at 300 K. Density functional theory calculations showed that the Pd atoms were activated by charge transfer from oxygen vacancies on the MgO substrate. It was quickly realized that this relatively simple model system could be ideal for understanding basic processes in catalysis[26]. A series of experimental and theoretical studies of MgO-supported single atoms followed [27-31] including the interesting observation that charge transfer could occur between adsorbed atoms (and clusters) and the underlying metal in ultrathin oxide film systems[32].

Meanwhile, Sykes and coworkers demonstrated that isolated single atoms could provide active sites for catalysis. Specifically, they embedded Pt group metals in less reactive metal hosts, creating so-called Single-Atom Alloys (SAA). The concept has proven successful, particularly for hydrogenation[33] and dehydrogenation reactions[34]. A comprehensive review was recently published[35]. Since Sykes has also contributed a perspective for this special issue of Surface Science highlighting the special properties of SAA[36], I will restrict the discussion here to SAC systems supported by non-metal substrates.

Aside from surface science studies, there was growing evidence from high-surface-area catalysis that oxide-supported single atoms could play a significant role in catalysis. For example, Flytzani-Stephanopoulos and coworkers[37] observed that the catalytic activity of a $Pt/CeO_2$ system in the water-gas shift reaction remained unchanged after metallic nanoparticles were removed, leading them to conclude that nonmetallic Pt species were solely responsible for the reactivity. In 2011, Zhang and coworkers[38] coined the term "single-atom catalysis" (SAC) and synthesized a catalyst composed exclusively of single Pt atoms supported on $FeO_x$—a challenging feat due to the tendency of Pt to sinter at high temperatures. They demonstrated that this catalyst outperformed both Pt and Au nanoparticle standards. This study, along with others, generated significant interest in the field (Ref [38] has received ~6000 citations at the time of writing). The synthesis approach, coupled with a suite of techniques—transmission electron microscopy (TEM), CO-diffuse reflectance infrared Fourier transform spectroscopy (CO-DRIFTS), and x-ray absorption spectroscopy (XAS)—used to verify atomic dispersion, has since become standard practice in SAC research.

Despite the growing popularity of SAC, it remains controversial due to difficulties in proving that single atoms are the true active sites; often, small clusters may be present or may form during reactions[39]. Another related issue is uncertainty surrounding the reaction mechanisms, partly due to limitations in current experimental techniques to characterize the active site directly. Computational modeling of SAC reactions typically relies on simplified systems, where the metal atom occupies an assumed position on a low-index termination of the support. However, assumptions in these models can have significant consequences. For instance, in the widely cited $Pt_1/FeO_x$ study, the computational model was based on Pt adsorbed at a 3-fold hollow site on an O-terminated $\alpha\text{-}Fe_2O_3(001)\text{-}(1\times1)$ surface. This model is reasonable at first glance, given the alignment of Pt atoms with Fe cations in TEM images of the as-synthesized catalyst[38]. However, the "x" in $FeO_x$ acknowledges that the reducing conditions in the catalyst activation step likely render the $\alpha\text{-}Fe_2O_3$ surface non-stoichiometric, an aspect that is not clearly resolved by experimental characterization. Furthermore, the nanoparticle support surface is heterogeneous, with various nanofacets, making an idealized $\alpha\text{-}Fe_2O_3(0001)$ surface model an oversimplification. Notably, surface science studies have found little evidence of a stable $(1\times1)$ surface structure on $\alpha\text{-}Fe_2O_3(0001)$, especially in reducing conditions: such a surface would be polar and would gain significant energy by losing surface oxygen atoms[40]. Thus, it is unsurprising that a CO oxidation mechanism involving the extraction of surface oxygen, known as a Mars-van Krevelen (MvK) mechanism, appears favorable[38]. However, using a more realistic reduced termination is challenging. Surface science studies indeed show that α-$Fe_2O_3$(0001) forms regions with varied terminations under mild reducing conditions[10], and then adopts a long-range "bi-phase" structure [15, 41]. All of these surfaces' structures remain debated[42]. With further reduction, an $Fe_3O_4$(111)-like termination appears[42], though it is itself controversial[43] and prone to forming a long-range biphase-ordered structure in reducing conditions[44]. In summary, while the modeling of the $Pt_1$/α-$Fe_2O_3$(0001) system illustrates that an MvK mechanism could explain catalytic activity observed in CO oxidation experiments, there is no direct experimental link confirming the structure or mechanism. This critique applies to many SAC studies, as theoretical models often lack a strong connection to experimental data, muddying our understanding of SAC capabilities and limitations.

In this author's opinion, the lack of experimental structural data remains a critical issue in the SAC field. As computational methods and resources advance, their capabilities will continue to grow, leading to an increasing trend of screening potential catalysts. However, such work can be unreliable when based on flawed assumptions. The $FeO_x$-supported SAC systems described above again provides a case in point: Screening studies based a bulk-truncated α-$Fe_2O_3$(001)-(1×1) surface model predict that $Pd_1$ and $Ni_1$ could outperform $Pt_1$[45] if a MvK mechanism is assumed, while other metals (including Ti) are preferred if an associative mechanism is assumed[44]. Replacing Pt with Ti would be an exciting breakthrough, yet without experimental validation, it remains doubtful whether these findings are reliable. Machine learning approaches will likely drive further increases in such studies; however, it remains debatable whether such predictions can reliably guide experimental efforts.

At this juncture, the surface science approach offers invaluable insights. Surface science experiments closely mirror computational models in structure and environment: the single-crystal support aligns with periodic boundary conditions in theoretical calculations, and the ultrahigh vacuum (UHV) environment simplifies the problem considerably. This approach allows to determine the structure of the support and the adatom position with high accuracy, facilitating the parallel interpretation of structural and reactivity trends with computational predictions. Scanning probe microscopy is ideal for tracking the evolution of model catalysts upon reactant exposure, and methods for studying catalytic mechanisms are well-established in the field. One critical service would be to provide benchmark experimental spectra for the most important methods utilized to characterize powder-based SACs. This involves measuring XPS core-level binding energies, the vibrational stretch frequencies of adsorbed CO molecules, and reference XAS spectra for metal atoms on well-defined surfaces, aiding in the identification of active sites in real systems. Additionally, it will be essential to assess these systems after exposure to environmental gases like water, as under-coordinated surface metal atoms are likely to coordinate with these molecules—an aspect rarely considered in studies of powder-based SAC systems[46, 47].

Inspired by these studies, we recently examined how water exposure impacts the stability of model SACs. We found that water stabilizes Rh on α-$Fe_2O_3$(012)[48], but has a neutral or destabilizing effect on various metals on rutile-$TiO_2$(110)[49] and anatase-$TiO_2$(101)[50]. The key difference lies in the binding structure: OH groups from the water can complete a square planar configuration for Rh atoms on α-$Fe_2O_3$(012)[48], enhancing stability. In contrast, no similar stabilizing structure is possible on the titania surfaces. These studies again underscore the crucial role of local environment in determining the behavior of single-atom catalysts.

A key finding from SAC research is that single atoms exhibit distinct catalytic properties compared to bulk or nanoparticle forms of the same metal. This difference arises because single atoms form chemical bonds with the support, often becoming ionic, which impacts reactant adsorption energies and catalytic activity. Our group highlighted this phenomenon through a systematic study of CO adsorption on Cu, Ag, Au, Ni, Pd, Pt, Rh and Ir atoms[51] supported on $Fe_3O_4(001)$. Using temperature-programmed desorption (TPD), we showed that CO adsorption energies at single-atom sites are significantly higher than on low-index extended metal surfaces (see Figure 1). Furthermore, the local coordination environment of the metal atom substantially affects adsorption energy. For example, CO binds only weakly at 5-fold Ni sites on $Fe_3O_4(001)$, where the physical and electronic structure strongly resembles NiO(001). On the contrary, CO desorbs with approximately 1.2 eV at the 2-fold coordinated adatom considered in Figure 1. This trend extends to other adsorbates: ethylene binds significantly more strongly at a 2-fold Rh adatom than at a 5-fold Rh adatom[52]. In these studies, the geometry of the metal adatom was determined through a combination of scanning probe microscopy and photoelectron spectroscopy, providing a robust basis for computational modeling.

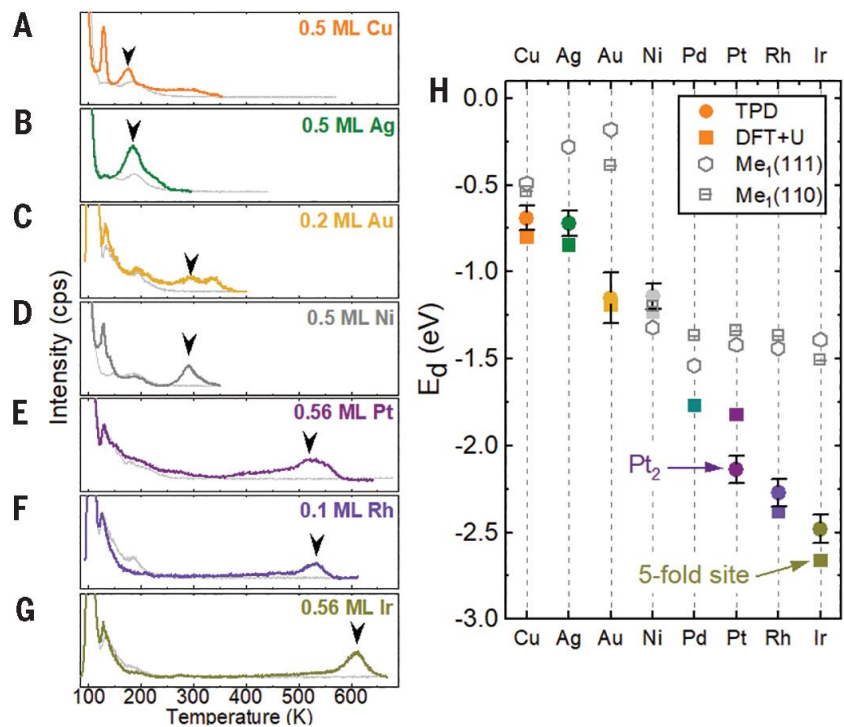

**Figure 1**. CO adsorption energy determination on $Me_1/Fe_3O_4(001)$ model catalysts using temperature-programmed desorption (TPD). Panels (A–G) show TPD curves for CO on different metal adatoms, where 1 monolayer (ML) represents one metal atom per surface unit cell ($1.42 \times 10^{14}$ atoms per square centimeter). Desorption peaks attributed to the metal adatoms are indicated by arrows. Light-gray curves depict CO TPD data from the ultrahigh-vacuum-prepared $Fe_3O_4(001)$ surface prior to metal deposition. Panel (H) presents a comparison of experimental and theoretical CO adsorption energies ($E_d$), as well as desorption energies, alongside literature values for metal (111) and (110) surfaces. Figure reproduced from Ref [51].

The importance of the coordination was also highlighted by a series of studies of the $Pt_1/CeO_2(111)$ system by Matolin, Libuda and coworkers[53-58]. $CeO_2(111)$ forms (100)-like step edges, creating a 4-fold coordination environment that stabilizes Pt as $Pt^{2+}$ ions, similar to the coordination observed in bulk PtO. However, this configuration limits reactivity, as the coordinatively saturated $Pt^{2+}$ cations do not interact strongly with adsorbates. Balancing stability and reactivity is crucial; ideally, one aims to achieve a configuration stable enough to prevent thermal sintering but under-coordinated enough to adsorb reactants effectively. In this context, we recently found that partially charged $Pt^{\delta+}$ cations at terrace sites on the $\alpha$-$Fe_2O_3(012)$ surface remain stable at room temperature due to a linear 2-fold coordination[59]. This finding is notable for two reasons. First, creating this adsorption geometry requires significant structural rearrangement of the support surface, involving the breaking of three Fe-O bonds. Second, this adatom geometry was identified using an automated search algorithm and was 0.85 eV more stable than a simple geometry based on an unmodified support. Such searches are arguably essential for SAC studies, as determining the minimum-energy adatom geometry is a pre-requisite to accurate system modeling.

In addition to iron oxides and ceria, several other metal oxides are commonly utilized as supports for SAC. These include materials where the structures of the low index facets are already known (anatase and rutile $TiO_2$, $SnO_2$, ZnO), so there is ample room for future study by surface scientists. The breadth of reactions catalysed by such systems is significant, and includes CO oxidation, the water gas shift reaction, and hydrogenation reactions. A particularly interesting application is hydroformylation, in which alkenes react with syngas to create aldehydes. Typically, hydroformylation is catalyzed by metal-organic Rh complexes dissolved in the liquid phase, posing challenges for product separation and catalyst reuse. This has sparked interest in a heterogeneous approach using solid catalysts. While Rh nanoparticles are active in this reaction, they generally exhibit low selectivity. Remarkably, several groups have reported that Rh SACs on various metal oxide supports (e.g., CoO, ZnO, $SnO_2$, $Al_2O_3$, $CeO_2$, $ReO_x$) achieve similar levels of selectivity as the homogeneous catalysts[60-64]. It is assumed that the reaction proceeds analogously, meaning that alkene, CO, and $H_2$ must be simultaneously adsorbed at the single-atom site at some stage. Such coadsorption at single-atom sites has yet to be directly observed but could be explored through surface science experiments. Such findings would reinforce the notion that SACs can function similarly to homogeneous catalysts. However, coadsorption of three reactants likely requires a low coordination environment for the Rh atom, as it is challenging to accommodate more than six ligands (including metal-support bonds). It is possible that Rh changes site from a stable resting position to a new configuration as reactants begin to adsorb, which could be tracked using surface science techniques.

Notably, many of the metal oxides listed above are semiconductors used in photocatalysis. Often, the oxide serves as the photoabsorber, while reaction kinetics are enhanced by a metal co-catalyst. It has been shown that adsorbed single atoms can also perform this function – so-called single-atom photocatalysis[65-67]. The mechanisms behind this enhancement are not fully understood, but it is possible that single atoms provide active sites not present on the oxide alone, increase the density of existing active sites, alter the electronic structure of the semiconductor surface, or passivate "trap sites." Given our understanding of metal oxide surface structures, surface science experiments could potentially rule some of these effects in or out, clarifying the role of single atoms in photocatalysis. Once determined experimentally, the relatively simple structure of the active site could provide an ideal model system for theoretical studies that include excited states in photocatalytic modeling.

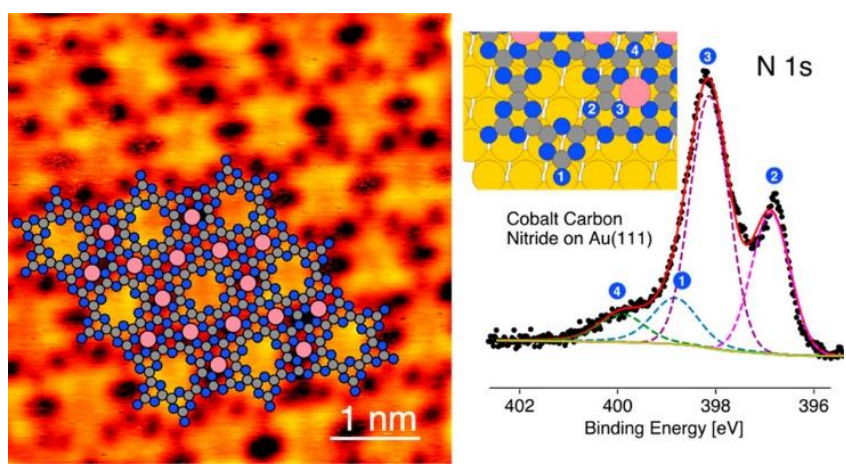

**Figure 2.** STM image of cobalt carbon nitride on Au(111), with a proposed model structure overlayed in which the cobalt has 4-fold coordination to nitrogen. XPS spectra (right) of the monolayer exhibit 4 different N species consistent with the model. Figure reproduced from Ref [68].

Moving away from metal oxides, carbon nitride (CN) is among the most prominent SAC supports[69-72]. Numerous studies have shown that metal atoms coordinated to nitrogen atoms on carbon nitride surfaces act as active catalysts, particularly for energy-related reactions[72] such as the oxygen evolution reaction (OER)[73], oxygen reduction reaction (ORR)[73], hydrogen evolution reaction (HER) [74], and nitrogen reduction reaction (NRR)[75]. CN is also widely used in photocatalysis[70]. However, a major limitation of these studies is the difficulty in experimentally determining the precise structure of carbon nitride[71, 75]. Computational models often assume a graphitic carbon nitride (g-CN) structure[73, 75], but this assumption is challenging to reconcile with the 4-fold nitrogen coordination seen in XAS experiments[76]. In recent years, several groups have studied the adsorption of melamine and melem—precursors in the formation of CN—using surface science techniques, finding that they form ordered molecular adlayers[77-80]. Thermal polymerization of these layers into carbon nitride structures can be achieved if the substrate (often a metal) is sufficiently reactive, although the presence of the metal may affect the reactivity of the resulting adlayer. For example, on Cu(111), Cu atoms appear to be incorporated into the CN structure upon heating[81]. Lauritsen and coworkers[68] polymerized melamine on Au(111) but found that co-depositing Co, Ni, or Mn was necessary, as melamine desorbed before thermal polymerization could occur. This process resulted in an ordered polymerized structure with single-atom sites for the chosen metal, although the structure differs from g-CN, as shown in Figure 2. Reactivity tests on this surface for OER and ORR in a custom-designed electrochemical cell revealed $CoO_x$ formation under OER conditions, with ORR performance dominated by the underlying Au(111). Nonetheless, this work highlights the potential of *in situ* preparation to yield model systems for studying CN-based SACs. A notable recent approach involved an adapted melem-like precursor with reactive 2,5,8-Triazido-s-heptazine entities, which could be removed thermally or through photo-illumination[82]. Although the structure of the polymerized layers on Au(111) and the (0001) facet of highly ordered pyrolytic graphite (HOPG) remains uncertain, it is not representative of g-CN as assumed in most CN-based catalyst models. If these structures could be determined experimentally, surface science techniques could be used to investigate atomic coordination, reactivity, and stability under ambient pressure and electrochemical conditions. Additionally, TPD could help benchmark computational approaches. It would be interesting if dihydride and di-oxide species could be observed, as these are predicted to be intermediates during HER and OER on single atom catalysts[83, 84].

Metal-organic frameworks (MOFs) represent a promising material class related to single-atom catalysis[85]. In MOFs, single atoms are coordinated by organic linker ligands, creating well-defined active sites. In this respect, MOFs closely resemble homogeneous catalysts, as their metal site properties can be tuned by varying linkers,[86] while sufficiently large linkers yield a porous 3D structure. MOFs can be mounted on surfaces, yielding so-called SURMOFs[87, 88]. Surface scientists have been studying on surface self-assembly of MOFs for a number of years[89, 90], and 2D MOFs[84, 91, 92] and covalent organic networks COFs[93-95] have been synthesized and tested for catalytic activity. A particularly nice, very recent example is shown in Figure 3, where a beautifully ordered Fe-TCNQ MOF was synthesized on a graphene layer on Ir(111)[96]. As with other SACs, a key question is the extent to which the reactivity of MOFs can be both homogeneous and tunable. Screening studies are beginning to emerge in the literature[97, 98]; however, reliable predictions will depend on adequate benchmark experimental data to provide a basis for comparison. Experimental benchmarking remains essential to validate computational predictions and guide future research in MOF-based SACs.

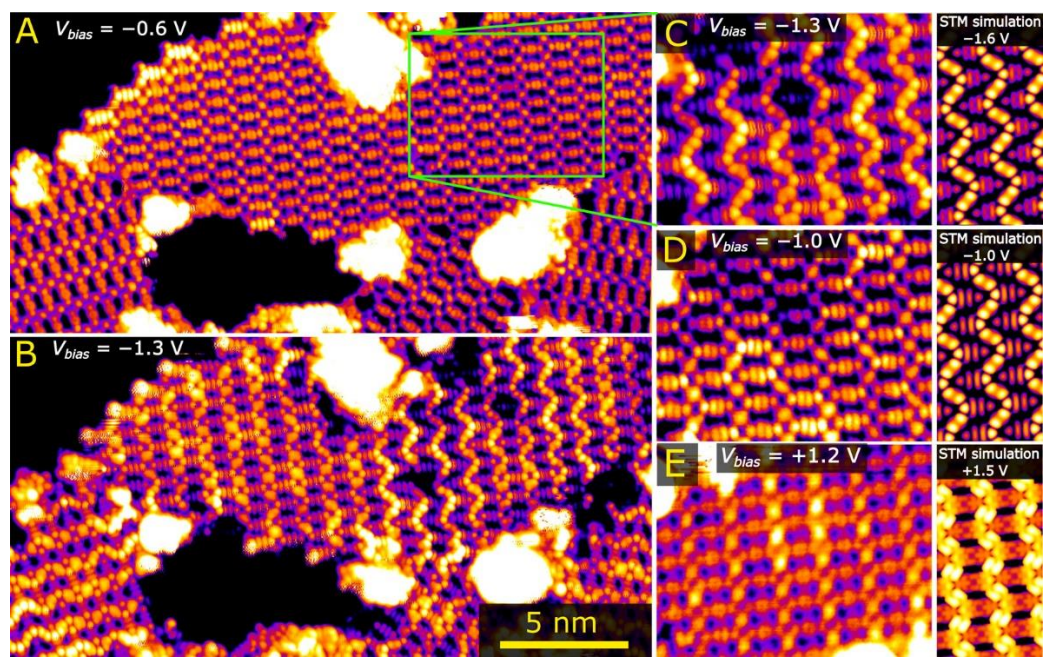

**Figure 3.** Room-temperature scanning tunneling microscopy (STM) images and computational models of the graphene-supported Fe-TCNQ 2D metal-organic framework (MOF) at varying sample biases. (A) At a sample bias of V = −0.6 V, the Fe-TCNQ structure appears planar and uniform. (B) A distinctive zigzag pattern emerges at a sample bias of V = −1.3 V. (C–E) High-resolution STM images of the same region at various biases, accompanied by STM simulations for comparison. Figure reproduced with permission from Ref. [96]. Copyright American Chemical Society 2024.

In summary, this article highlights the urgent need for reliable benchmark data in the field of SAC. Surface science is ideally suited to provide this data due to its capacity to precisely control and characterize the structure of active sites. While close integration with theoretical computations is vital, it is equally important to validate findings from UHV-based studies in realistic conditions. Fortunately, the ongoing development of *operando* techniques—such as NAP-STM, NAP-XPS, and PM-IRAS, along with their electrochemical variants—positions surface science to effectively bridge the pressure gap. These advancements provide a robust platform for observing catalytic behavior under conditions closer to actual operating environments, supporting the continued development of this promising technology.

**Acknowledgments**

Funding from the European Research Council (ERC) under the European Union's Horizon 2020 research and innovation program (grant agreement No. [864628], Consolidator Research Grant 'E-SAC') is acknowledged. This research was funded in whole or in part by the Austrian Science Fund (FWF) 10.55776/F81 and 10.55776/COE5. For open access purposes, the author has applied a CC-BY public copyright license to any author-accepted manuscript version arising from this submission. I acknowledge Prof. Ulrike Diebold (TU Wien) for critically reading the manuscript prior to submission.